# A Fast Multiple Attractor Cellular Automata with Modified Clonal Classifier for Splicing Site Prediction in Human Genome

Pokkuluri Kiran Sree [1], Inampudi Ramesh Babu [2], SSSN Usha Devi N [3]

1. Dept of CSE, JNTU Hyderabad, profkiransree@gmail.com
2. Dept of CSE, ANU, Guntur, drirameshbabu@gmail.com
3. Dept of CSE, University College of Engineering, JNTUK.

**Abstract**

Bioinformatics encompass storing, analyzing and interpreting the biological data. Most of the challenges for Machine Learning methods like Cellular Automata is to furnish the functional information with the corresponding biological sequences. In eukaryotes DNA is divided into introns and exons. The introns will be removed to make the coding region by a process called splicing. By indentifying a splice site we can easily specify the DNA sequence category (Donor/Accepter/Neither).Splicing sites play an important role in understanding the genes. A class of CA which can handle fuzzy logic is employed with modified clonal algorithm is proposed to identify the splicing site. This classifier is tested with Irvine Primate Splice Junction Database. It is compared with NNspIICE, GENIO, HSPL and SPIICE VIEW. The reported accuracy and efficiency of prediction is quite promising.

## 1. Introduction

DNA (Horwitz,1998), of the organism is organized as one or more chromosomes. The total sequence is called as genome. Genome contains genes which encode proteins. Genes can be found in forward as well as reverse strand. Gene holds the coded data important to make a protein. Most of the work in our cells and body will be done by proteins. Human genome will have 30,000 to 40,000 genes. The function of certain genes is to control the expression (turning on or off the production) of other genes, hence forming networks of relationships and interactions. DNA sequence determines Protein Sequence. Protein Sequence determines protein structure. Protein structure determines protein function.

Gene is a particular sequence of DNA that consist information to formulate a full or part of protein or a RNA molecule. Genes also determines the characteristics of the particular organism. These are passed from ancestors to offspring, causing some of the offspring to inherit the characteristics of their ancestors.

Prokaryotes are single cell organisms which are very simple and primitive. It is very simple compared with eukaryote which lacks a membrane bound nucleus. Major viruses and bacteria will be belonging to this category. Eukaryotes are multi cell organisms which are the complex forms of prokaryotes. Eukaryotes will consist of nucleus which will contain the major genetic features stored in it. Most of the large organisms are eukaryotes which includes human, animals and plants. In eukaryotes slicing is the process of removing Intorns as show in fig 1.

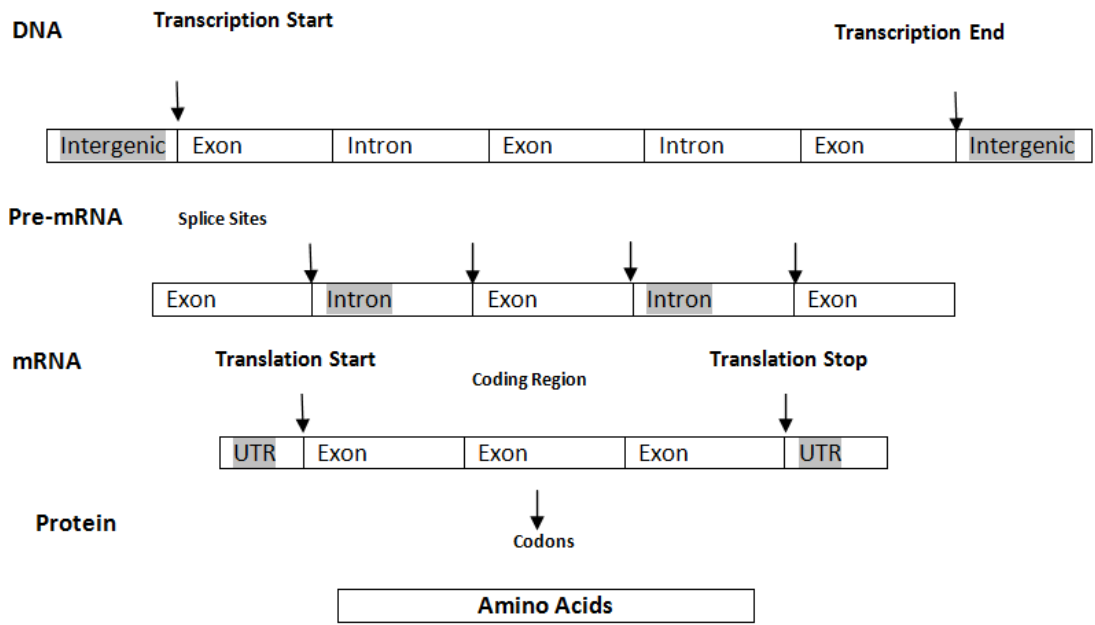

Figure 1: Transcription, Translation and Splicing

Von Neumann and Stanislaw Ulam (Von Neumann, 1996) initially proposed the model of Cellular Automata in 1940. Stephen Wolfram , (Wolfram, 1994) did a detailed study on one-dimensional CA (Elementary CA). He later published a book on "A New Kind of Science" in 2002 which dealt with basic and neighborhood structure of CA has pulled in scientists from different disciplines. It has been subjected to thorough numerical and physical dissection for most recent fifty years and its requisition has been proposed in diverse extensions of science - both social and physics.

## 2. Design of MACA based Modified Clonal Classifier

A Cellular Automata which uses fuzzy logic is an array of cells arranged in linear fashion evolving with time. Every cell of this array assumes a rational value in the interval of zero and one. All this cells changes their states according to the local evaluation function which is a function of its state and its neighboring states.

A Cellular Automata Maji(2003) which uses fuzzy logic is an array of cells arranged in linear fashion evolving with time. Every cell of this array assumes a rational value in the interval of zero and one. All this cells changes their states according to the local evaluation function which is a function of its state and its neighboring states. The synchronous application of the local rules to all the cells of array will depict the global evolution.

Assume n represent the number of fuzzy states Nedunuri(2013), Sree(2014), and qj denotes the fuzzy state in figure 2, a rational value in between zero and one will assigned to each state.

$Q_j = j/n-1$ whre j=1.1.2………n-1.

The algorithms takes input as DNA sequence and the maximum population and give output as the class, matrix representation and rule specification.

Input: S = {S1, S2, · · · , Sl}, Training Set, Maximum Population Mmax).

Output: Matrix Representation T, F, and information of the class

Begin

Step 1: Generate 500 new chromosomes for Initial Population.

Step 2: Initialize Maximum Population MM=zero; PP← IP.

Step 3: Compute fitness FF for each chromosome of PP

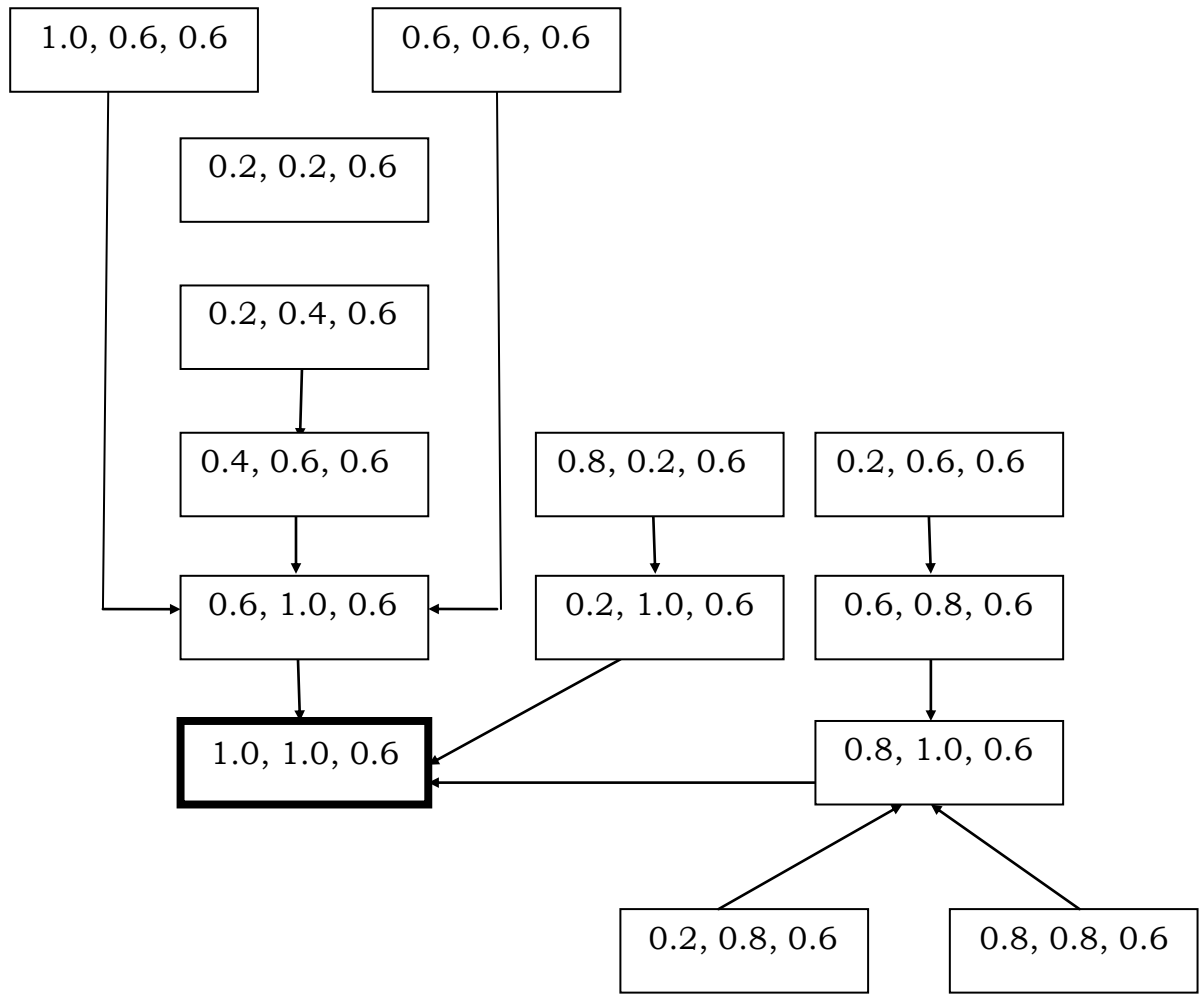

Figure 2: MACA-Modified Clonal Classifier Tree with basin 1, 1, 0.6

Step 4: Store T, F, and corresponding class information for which the fitness value FF = 1.

Step 5: If FF = 1 for at least one chromosome of PP, then go to Stop.

Step 5a: Donor Module

Step 5b: Accepter Module

Step 6a: Construct the MACA-CC tree based on 5a,5b

Step 6: Order chromosomes in order of fitness.

Step 7: Increment Maximum Population (MM).

Step 8: If GC > Gmax then go to Step 11.

Step 9: Form NP by operations of Modified Clonal algorithm

Step 10: PP← NP; go to Step 3.

Step 11: Out Put and Store T, F, and corresponding class information for which the fitness value is maximum.

Step 12: Stop.

(Baten, 2006),(Pertea, 2001),(Maji 2014) has developed algorithms for predicting splicing sites.

## 3. Experimental Results

The data sets are taken from Irvine Primate Splice Junction Database[9]. It is compared with NNspIICE, GENIO, HSPL and SPIICE VIEW. The predictive accuracy reported is 94.3% show in table 1.

### 4.1 Parameters for testing splicing sites

The important statistics to look at include:
1. True Positives (TP1): Number of correctly predicted acceptors.
2. False Positives (FP1): Number of incorrectly predicted donors.
3. True Positives (TP2): Number of correctly predicted acceptors.

4. False Positives (FP2): Number of incorrectly predicted donors.

5. True Negatives (TN1): Number of correctly predicted non acceptors

6. False Negatives (FN1): Number of incorrectly predicted non donors

7. True Negatives (TN2): Number of correctly predicted non acceptors

8. False Negatives (FN2): Number of incorrectly predicted non donors

Using the above measures following are calculated for two sets.

- Actual Positives (AP) = TP + FN
- Actual Negatives (AN) = TN + FP
- Predicted Positives (PP) = TP + FP
- Predicted Negatives (PN) = TN + FN
- Sensitivity (SN) = TP / (TP + FN)
- Specificity (SP) = TP / (TP + FP)

Table 1: Sensitivity and Specificity Reporting

| Algorithm/ Coding Measure | Sensitivity | Specificity |
| --- | --- | --- |
| NNspIICE | 66.3 | 67.4 |
| GENIO | 69.36 | 72.2 |
| HSPL | 73.3 | 76.5 |
| SPIICE VIEW | 82.3 | 84.3 |
| MACA-MCC | 88.6 | 90.3 |

### Donor site predictions

| Start | End | Score | Exon | Intron |
|---|---|---|---|---|
| 114 | 128 | 0.45 | gcaactggtgtgtcg | |

| Start | End | Score | Exon | Intron |
|---|---|---|---|---|
| 174 | 188 | 0.45 | gcaactggtgtgtcg | |

Direct chain.
 Acceptor(AG) sites. Treshold    4.175 (90%).
    1 P:    783 W: 4.17 Seq: tctgaAGgacag
    2 P:    814 W: 4.65 Seq: gtcttAGacatc
 Donor(GT) sites. Treshold    6.099 (90%).
    1 P:    181 W: 7.36 Seq: aactgGTgtgtc
    2 P:    451 W: 6.66 Seq: ttttgGTgggtc

### Acceptor site predictions

| Start | End | Score | Intron | Exon |
|---|---|---|---|---|
| 703 | 743 | 0.65 | atcacctctccatctctgaaggacaggattcactgtgtggc | |

| Start | End | Score | Intron | Exon |
|---|---|---|---|---|
| 763 | 803 | 0.65 | atcacctctccatctctgaaggacaggattcactgtgtggc | |

Reverse chain.
 Acceptor(AG) sites. Treshold    4.175 (90%).
    1 P:    359 W: 4.90 Seq: attttAGagtag
    2 P:    423 W: 7.50 Seq: ccttcAGagatg
    3 P:    561 W: 4.45 Seq: catctAGcccca
    4 P:    620 W: 7.00 Seq: ttaacAGaatat
    5 P:    864 W: 9.30 Seq: cgtctAGgttat
    6 P:    963 W: 6.05 Seq: tatctAGataaa
 Donor(GT) sites. Treshold    6.099 (90%).
    1 P:   1140 W: 6.10 Seq: ttcacGTaaatt
    2 P:   1192 W: 8.90 Seq: tcaatGTaagcc

### 4. Conclusion

We have successfully developed and tested the MACA based modified Clonal Classifier for splicing sites in eukaryotes particularly humans. The proposed classifier is tested for specificity and sensitivity. It is compared with important splicing programs available. The results obtained are found promising and comparable. This classifier is also observed and tested for the amount of time it will be taking to predict the splicing site was found as .02 ms A sensitivity of 84.5% and specificity of 92.7 were reported.